\documentclass[12pt]{iopart}

\usepackage{graphicx}
\usepackage[utf8]{inputenc}
\usepackage{amssymb}
\usepackage{caption}
\usepackage[sort, compress]{cite}

\begin{document}
%%PRIDANA UPRAVA DOCUMENTU
%\pagestyle{empty}

\title[Self-organization phenomena in cold atmospheric plasma slit jet]{Self-organization phenomena in cold atmospheric pressure plasma slit jet}

\author{K Polášková$^{1,2,3}$, D Nečas$^{2,3}$, L Dostál$^2$, M Klíma$^2$, P Fiala$^2$ and L. Zajíčková$^{1,2,3}$}

\address{$^1$ Department of Condensed Matter Physics, Faculty of Science, Masaryk University, Kotlářská 2, CZ-61137 Brno, Czech Republic}
\address{$^2$ Faculty of Electrical Engineering and Communication, Brno University of Technology, Technická 12, CZ-61600 Brno, Czech Republic}
\address{$^3$ Plasma Technologies, Central European Institute of Technology - CEITEC, Brno University of Technology, Purkyňova 123, CZ-61200 Brno, Czech Republic}
\ead{lenkaz@physics.muni.cz}

\begin{abstract}
The RF plasma slit jet, which produces 150\,mm wide streaming plasma outside the jet body, exhibits exciting self-organization phenomena that resemble the self-organized patterns of dielectric barrier discharge (DBD) filaments. Similarly, as in DBD, the filaments are surrounded by an inhibition zone that does not allow two filaments to come closer to each other. With fast camera imaging, we observed the filamentary character of the discharge in all the studied gas feeds (Ar, Ar/N$_2$, and Ar/O$_2$). Still, the visual appearance of the filaments in the plasma and their interaction with a dielectric surface depended significantly on the gas feed. As the breakdown voltage in pure Ar is relatively low compared to the applied one, new filaments form frequently. Such newly created filaments disrupted the characteristic inter-filament distance, forcing the system to rearrange. The frequent ignition and decay processes in Ar led to short filament lifetimes (0.020--0.035\,s) and their high jitter speed (0.9--1.7\,m/s), as determined with an image processing custom code based on Gwyddion libraries. The number of filaments was lower in the Ar/O$_2$ and Ar/N$_2$ mixtures. It was attributed to a loss of energy in the excitation of rotational and vibrational levels and oxygen electronegativity. Since the probability of low-current side discharges transitioning into the full plasma filaments was limited in the gas mixtures, the self-organized pattern was seldom disrupted, leading to lesser movement and longer lifetimes. Unlike in Ar or Ar/O$_2$, the constricted filaments in Ar/N$_2$ were surrounded by diffuse plasma plumes, likely connected to the presence of long-lived nitrogen species. We demonstrated in the polypropylene treatment that the self-organization phenomena affected the treatment uniformity.

\end{abstract}


\submitto{\PSST}
\maketitle

% \linenumbers

\section{Introduction}
\label{sec1}
Until recently, plasma enhanced chemical vapor deposition (PECVD) was predominantly carried out in low, rather than atmospheric, pressure systems because of better control over the process (thanks to less pronounced effect of gas dynamics and diffuse nature of low pressure discharges), and better film quality as assessed from the view of film uniformity and employment of energetic ions. However, atmospheric pressure discharges offer unique possibilities when creating local structures and utilizing liquid precursors including biomolecules or drugs~\cite{schafer2017, boscher2016, porto2018}, which are all great advantages in the age of nanotechnologies and biotechnologies. Therefore, the research of cold atmospheric pressure (CAP) discharges and related plasma-chemical processes became one of the leading topics in low-temperature plasma physics. 
%PROPOSAL CITATIONS
%proposal1 = Schafer et al. (2017) => Liquid assisted plasma enhanced chemical vapour deposition with a non-thermal plasma jet at atmospheric pressure
%proposal2 = Boscher et al. (2016) => Liquid-assisted plasma-enhanced chemical vapor deposition of a-cyclodexterin/pdms composite thin film for the preparation of interferometric sensors
%proposal3 = Porto et al. (2018) => On the plasma deposition of vancomycin-containing nano-capsules for drug delivery applications

%Due to high collision rates, a growing electron avalanche can generate appreciable charge density at its tip after traveling a short distance. The local field caused by the charge separation resulting from the difference in drift velocities of electrons and ions is superimposed on the applied field. Collisional ionization in the high-field region at the streamer head leads to fast propagation of the ionization region and the formation of a bright plasma filament~\cite{Kogelschatz02}. 
Two typical configurations of CAP discharges are planar dielectric barrier discharges (DBD) and capillary plasma jets. They tend to have filamentary character due to the streamer type of breakdown. Although DBDs can be homogeneous in some special cases such as 
atmospheric pressure glow discharge (APGD)~\cite{massines2003,massines2012} and atmospheric pressure Towsend discharge (APTD)~\cite{massines2012,elias2015}, 
%M. Eliáš et al. JAP 117 (2015) 103301 
%F. Massines et al. SCT 174-175 (2003) 8 ~\cite{massines2003}
% Massines et al. Plasma Process. Polym. 2012, 9, 1041–1073  DOI: 10.1002/ppap.201200029
in a common DBD, individual filaments (microdischarges) compete for the available surface area of the dielectric to deposit their charge patterns. Since they encounter residual charges accumulated on the dielectrics from previous discharge phases, the memory effect is a dominant feature. There may be additional memory effects due to residual gas heating and the presence of charged and excited species left over from the previous half period, if the repetition frequency is high enough~\cite{kogelschatz2002}. %IEEE TRANSACTIONS ON PLASMA SCIENCE, VOL. 30, NO. 4, AUGUST 2002

% self-organization in DBD
In a confined geometry, microdischarges with their surface discharges can be quite regularly arranged due to the memory effects. Guikema {\it et\,al.}~\cite{guikema2000} %Walhout’s group 
demonstrated array of regularly spaced microdischarges in atmospheric pressure He/Ar one-dimensional DBD.  
%[74] J. Guikema, N. Miller, J. Niehof, M. Klein, and M. Walhout, Phys. Rev. Lett. 85, 3817 (2000).
Regular patterns have also been observed in 2-D barrier discharges for various gases and configurations~\cite{boyers1982,breazeal1995,muller1999,shirafuji2003}. 
%Boyers and Tiller~\cite{} published a paper on plasma-bubble domains obtained in atmospheric-pressure helium confined between large-area closely spaced plane-parallel dielectric-covered electrodes. 
%I. Müller, C. Punset, E. Ammelt, H.-G. Purwins, and J.-P. Boeuf, “Self-organized filaments in dielectric barrier glow discharges,” IEEE Trans. Plasma Sci., vol. 27, pp. 20–21, Feb. 1999.
%W. Breazeal, K. M. Flynn, and E. G. Gwinn, “Static and dynamic two-dimensional patterns in self-extinguishing discharge avalanches,” Phys.Rev. E, Stat. Phys. Plasmas Fluids Relat. Interdiscp. Top., vol. 52, pp.1503–1515, Aug. 1995.
% https://aip.scitation.org/doi/pdf/10.1063/1.1613796
Boeuf {\it et\,al.}~\cite{Boeuf2012} created a self-consistent fluid discharge model of a quasi-1D DBD, similar to the one of Guikema {\it et\,al.}~\cite{guikema2000}, to simulate the complex self-organization phenomena in DBD over long time scales.
Combining the simulation and fast camera diagnostics, they proposed clear physical description and explanation of the mechanisms responsible for the generation, annihilation, motion and self-organization of discharge filaments in
DBDs. They concluded that these mechanisms were triggered by low-current “side discharges” generated during the same half-cycle in the vicinity of an isolated filament beyond the inhibition zone associated with charge spreading along the dielectric surface.
%play an essential role in the triggering of these mechanisms.
% Boeuf2012 https://doi.org/10.1063/1.4729767

%% self-organization in plasma jets
The self-organization phenomena were also observed in cold plasma jets working at the atmospheric pressure. Sch\"afer {\it et\,al.}~\cite{Schafer09}
%doi:10.1088/0741-3335/51/12/124045 Plasma Phys. Control. Fusion 51 (2009) 124045
reported on the self–organization effects in a dielectric barrier plasma jet driven by 27.12 MHz radio frequency (RF) generator. The dielectrics was a quartz capillary (4 mm in diameter) around which the RF and grounded electrodes were placed. They observed several regularly repeated arrangements of plasma filaments in which the number of filaments and the stationary or rotating (locked) modes changed depending on the plasma conditions. The regular filament patterns observed in the horizontal projection of the RF plasma jet in the stationary mode resemble the equidistant arrangements of filaments in the DBDs described above. %tohle nesouhlasi s tim, co pise sam Honza v clanku, kde se tvari ze je to spis locked mode co pripomina ty usporadane filamenty v DBD. To je ale blbost. Zvlast kdyz cituje [22] ktera neni o DBD.

We investigated self-organization phenomena in the RF plasma slit jet that produces wide streaming plasma outside the jet body~\cite{polaskova2021}. This type of atmospheric pressure discharge uses a different configuration than the parallel plate dielectric barrier discharges or capillary plasma jets discussed above. The coupling is achieved through an RF coil designed around the dielectric slit inserted in a tunable metal cavity. The plasma is ignited in argon flow to which other gases can be admixed. Thus, it provides large area plasma processing capability in the configuration with flexible distance from the processed surfaces. Fast camera imaging revealed exciting self-organization phenomena of the plasma filaments that resembles the self-organized patterns of DBD filaments. Similarly, as in DBD, the filaments are surrounded by an inhibition zone that does not allow two filaments to come closer to each other. The filamentary character of the discharge was observed with fast camera imaging for all the studied gas feeds (Ar, Ar/N$_2$, and Ar/O$_2$), but the visual appearance of the filaments in the plasma and their interaction with a dielectric surface depended significantly on the gas feed. We also demonstrated in the polypropylene treatment that the self-organization phenomena affected the treatment uniformity. 

\section{Experimental details}
\label{sec2}
\subsection{RF plasma slit jet and electromagnetic field simulation}
%{Construction and experimental conditions of RF plasma slit jet}

\begin{figure}[ptb]
 \includegraphics[width=1\textwidth]{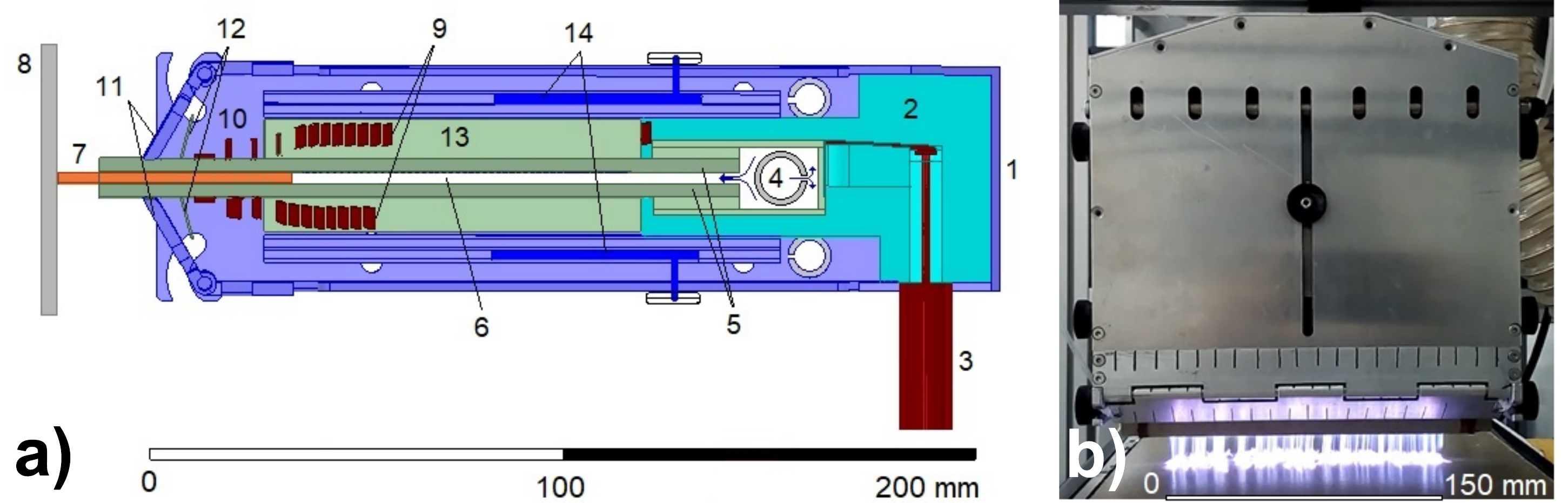}
 \centering
 \caption{(a) Schematic drawing of the RF plasma slit jet (side view): 1-- aluminum cover/shielding, 2 -- load-bearing element, 3 -- coaxial cable, 4 -- Ar flow homogenization region, 5 -- mica composite slit body, 6 -- slit, 7 -- plasma region, 8 -- sample, 9 -- resonance coil, 10 -- high-voltage radio-frequency electrodes, 11 -- system of movable grounded electrodes, 12 -- dielectric plates of an electrical breakdown limiter, 13 -- plates supporting the coil turns, 14 -- conductive plates of resonance matching circuit. (b) Front view of Ar plasma slit jet in contact with a mica composite substrate.}
 \label{fig:RF_plasma_slit_jet}
\end{figure}

The RF plasma slit jet (PSJ) configuration is shown in figure~\ref{fig:RF_plasma_slit_jet}. The mica-composite slit was placed inside a specially designed coil serving as the periodic deceleration structure. The jet body was situated inside the tunable cavity made of the fixed metal cover/shielding and movable conductive plates used as a resonance matching circuit. Therefore, the RF plasma slit jet could be connected directly to the RF generator (through a coaxial cable), and the set-up does not need a separate matching unit. We designed the PSJ to work at the frequency of 13.56\,MHz in the width of 150\,mm. The RF power of 500 or 600\,W was delivered by the CESAR 136 generator (Advanced Energy). 
The working gas was injected into the jet through a tube with evenly distributed holes facing the topmost metal cover. The set-up ensured random flow turbulence. As the gas started flowing downwards through the slit, it homogenized, leading to the uniform flow at the slit exit. %Hence, if gas mixtures are used, individual gases must be mixed before entering the RF plasma slit jet.
Three working gas feeds were studied: Ar, Ar/O$_2$ and Ar/N$_2$. Ar flow rate was set to either 67 or 100\,slm. O$_2$ flow rate was 1\,slm and the N$_2$ flow rate was set to 1.5\,slm, if not stated otherwise. The distance between the slit exit and the processed dielectric plate was kept constant at 10\,mm. 

The electromagnetic field was simulated using a wave-theory based numerical model formulated from the reduced Maxwell's equations. The model, implemented using Ansys EMAG platform, was conceived as a fully non-stationary problem defined on the basis of the telegrapher's equations. The differential equations were discretized using the finite element method and solved in a close-to-real three dimensional geometry of the PSJ positioned above the grounded plate covered with 3\,mm thick dielectric mica composite. %The plasma was approximated as a conductor (electrical conductivity 100\,S/m, permittivity and permeability equal to one) in a quasi-stationary state. 
The materials used in the PSJ body were characterized by tabulated constitutive parameters. 
%dávat zde tuto větu - KATKA, Lenka - je to dobry, aspon to vypada vedecteji :-)
As it was not always possible to find the values of constitutive parameters corresponding to MHz frequencies, they were approximated by the ones measured at the frequency closest to 13.56\,MHz. 
For simplicity, influence of heating on the material constants was neglected. The simulations were carried for power 600 W and the distance between the slit exit and the opposing dielectric plate 10\,mm.

%ANSYS-EMAG SW => Maxwell eq.
%měrná vodivost plazmatu 100 Siemens/m, permitivity 1, permeabilita 1
%přiblížení kvazistacionární stav => vodič
%výkon 600W, 13.56 MHz
%mřížka zhuštěná v oblasti plazmatu a cívky
%okolí: dielektrikum permitivity 1, permeabilita 1
%materiály trysky = tabelované konstanty
%a wave-theory based numerical model formulated from the reduced Maxwell equations
%the model was conceived as a fully non-stationary problem defined on the basis of the telegrapher's equations
%constitutive parameters = permitivita, permeabilita a vodivost

%%Obrázek
%%Obrázek

%%Obrázek
%%Obrázek

\subsection{Fast camera imaging of plasma slit jet}

Investigation of the filaments behavior and dynamics of their interaction with the dielectric surface (mica composite, thickness of 6\,mm) was realized by high-speed camera imaging using two synchronized ultrafast cameras, Photron FASTCAM SA-X2 (side view) and Olympus i-SPEED 726R (front view). The discharge behavior was monitored over a time interval of 1.09\,s at frame exposure time of 25\,\textmu s (40000 frames/second). The front-view 12bit greyscale video frames had resolution $2048\times190$ pixels with pixel size approximately 90\,\textmu m. The side-view video frame resolution was $512\times512$ pixels with a slightly smaller pixel size 65\,\textmu m.

Video data were processed using custom code based on Gwyddion~\cite{Necas2012} image processing libraries. Two regions were analyzed in each video, one where filaments formed and another covering the base region where plasma interacted with the substrate. Several parameters were evaluated from the reconstructed time dependencies of filament positions: mean and minimum inter-filament distance, distance upon formation and decay, `jitter' speed (movement speed as the shortest accessible time scale), mean filament lifetime and the rate at which the gap to the nearest neighbors expanded/contracted upon filament formation and decay. A more detailed description of the video data processing is in Supplementary Information (section~S2).

\subsection{Visualization of plasma-surface interaction by treatment of polypropylene}

The plasma treatment uniformity was visualized by treating polypropylene (PP) sheets (PP-H Natural, Omniplast), 5\,mm in thickness, placed on a mica composite substrate moved by a conveyor belt beneath the PSJ. %The sheets were produced by the extrusion of a PP granulate using three rollers. To prevent scratches a protective foil was used on both sides. The bottom side of the PP sheet that was in the contact with the rollers during the extrusion was rougher than the top side. 
The PP sheets were cut into 100$\times$25\,mm sized strips. For the experiments, the less rough side of PP was chosen after cleaning with isopropanol and cyclohexane as described previously~\cite{polaskova2021}.
The treatment uniformity was studied at the constant speed 100 or 250\,mm/s. 
%Standardly, one pass was used.

\begin{figure}[ptb]
 \includegraphics[width=0.9\textwidth]{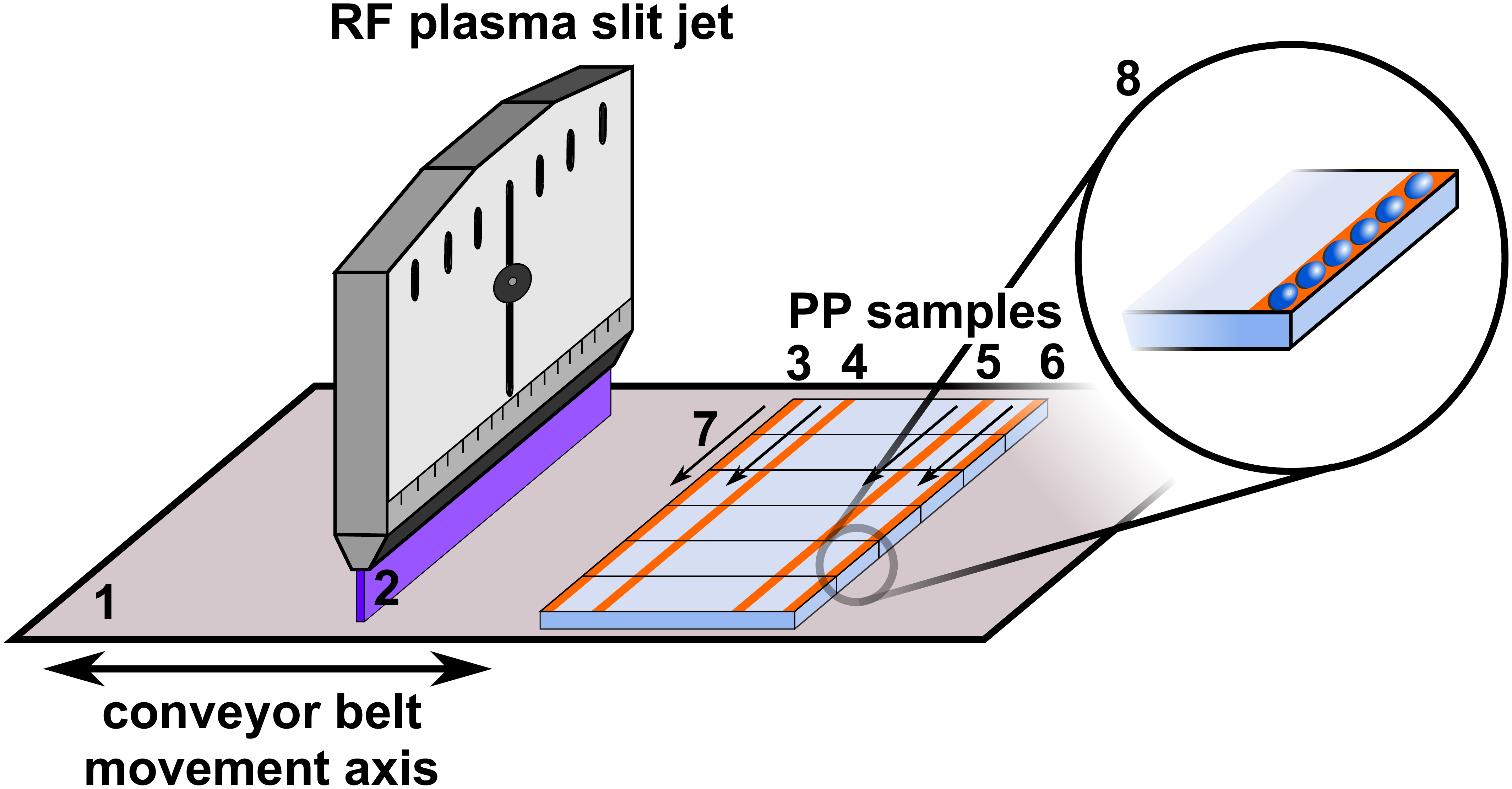}
 \centering
 \caption{Schematics of PP treatment set-up and the WCA uniformity measurement procedure: 1 -- mica composite, 2 -- plasma, 3 -- first line of the droplet deposited on the PP, 4 -- second line of the deposited droplet, 5 -- third line of the deposited droplet, 6 -- fourth line of the deposited droplet, 7 -- direction in which the droplets were deposited, 8 -- a line of sessile droplets deposited on one PP strip.}
 \label{fig:WCA-measurement}
\end{figure}

%\subsection{Water contact angle and ink test surface analyzes}
Water contact angle was measured by the sessile droplet method (3\,\textmu l of demineralized water) using the See System (Advex Instruments). Six PP strips placed one next to another with the shorter edges oriented parallel to the jet slit were used for one measurement~(figure~\ref{fig:WCA-measurement}). It consisted of depositing four lines of six droplets per one PP sample. Two droplet lines were placed near the short edges of the PP sample and the remaining two lines were both placed $\sim$20\,mm from the lines at the edges. Overall, 36 droplets per one line were deposited for all 6 strips. 
The influence of aging was ruled out using two sets of samples differing in the movement direction during the treatment while the order of deposited droplet lines remained the same. The reported WCAs were calculated as an average of at least six values determined using the three-point method implemented in the See System 7.0 software. The standard errors were computed for the 95\% confidence interval. 

The treated and untreated areas were visualized by the ink test Arcotest -- testing marker QuickTest 38 (TQC). The marker is designed to discriminate polymer surfaces with the free energy above and below 38\,mN\,m$^{-1}$. As the WCA measurement modified the PP surface, only the area between the 2$^\mathrm{nd}$ and the 3$^\mathrm{rd}$~(Figure~\ref{fig:WCA-measurement}) droplet line was used for the ink test analysis.

%%Obrázek
%%Obrázek

%%Obrázek
%%Obrázek

\section{Results and discussion}
\label{sec:results}
\subsection{Initial electromagnetic field and RF power coupling}
\label{sec:discharge_character}
% z GACR proposalu
%The models of RF discharges with capacitive coupling discuss the spatial distribution of the electric field E while the magnetic component is neglected. The RF inductive coupling (ICP -- inductively coupled plasma) utilizes a high-frequency EM field with prevailing intensity of the magnetic component (H typically reaches $10^3$--$10^4$\,A/m) and low intensity of the electric field E (typically $10^2$\,V/m). The models typically take into account the both EM components but not as a propagating EM wave. Contrary, the discharges at microwave frequencies or wave-heated discharges require the modelling of EM wave propagation and absorption. K. Gadonna et al. modelled the atmospheric pressure microwave torch using some input parameters (electron density and temperature, gas temperature) from the experiments and discussed the spatial distribution of the Poynting vector and the efficiency of the power absorption~\cite{proposal10}. Time- and space-resolved magnetic probe measurements in combination with measurements of the plasma parameters were carried in the helicon discharges and the Poynting vector and the absorbed power density were deduced~\cite{proposal11}.

CAP plasma jets are usually based on the capacitive coupling. Different configurations of electrodes and various excitation frequencies from kHz up to tens of MHz have been developed and investigated starting from the 90s when first configurations appeared in the patents~\cite{CZpatent98} or publications~\cite{Koinuma92,Janca_SCT99}. 
% Koinuma ... Appl. Phys. Lett. 60, 816 (1992); doi: 10.1063/1.106527
The various RF set-ups include the unipolar RF discharges (13.56\,MHz) 
%(also called barrier torch discharge, BTD) s kapilarou a prstencovou el.
with an inner hollow RF electrode~\cite{CZpatent98,Janca_SCT99,USpatent2013} or the outer ring RF electrode on a quartz capillary~\cite{Hubicka02}, %taky Klima06
%Z. Hubička et al. PSST 11 (2002) 195
the configuration with the central pin electrode in a dielectric capillary and outer grounded ring electrode (kINPen at about 1\,MHz~\cite{gaborit2014,reuter2018}), two ring electrodes on the dielectric capillary~\cite{Schafer09}, 
%J. Schäfer et al. PPCF 51 (2009) 124045
and the COST reference microplasma jet with parallel RF (13.56\,MHz) and grounded electrodes~\cite{beijer2016,liu2021}. %He
%S. Reuter et al. JPD 51 (2018) 233001 
The properties of high frequency discharges strongly depend on the type of plasma excitation. The RF plasma slit jet (PSJ)~\cite{polaskova2021} uses a different coupling principle than the jets described above. The coupling is achieved through special elements integrated in the plasma jet, periodic deceleration structures, consisting of combinations of inductors with specifically designed varying geometry and winding.

\begin{figure}[ptb]
 \includegraphics[width=1\textwidth]{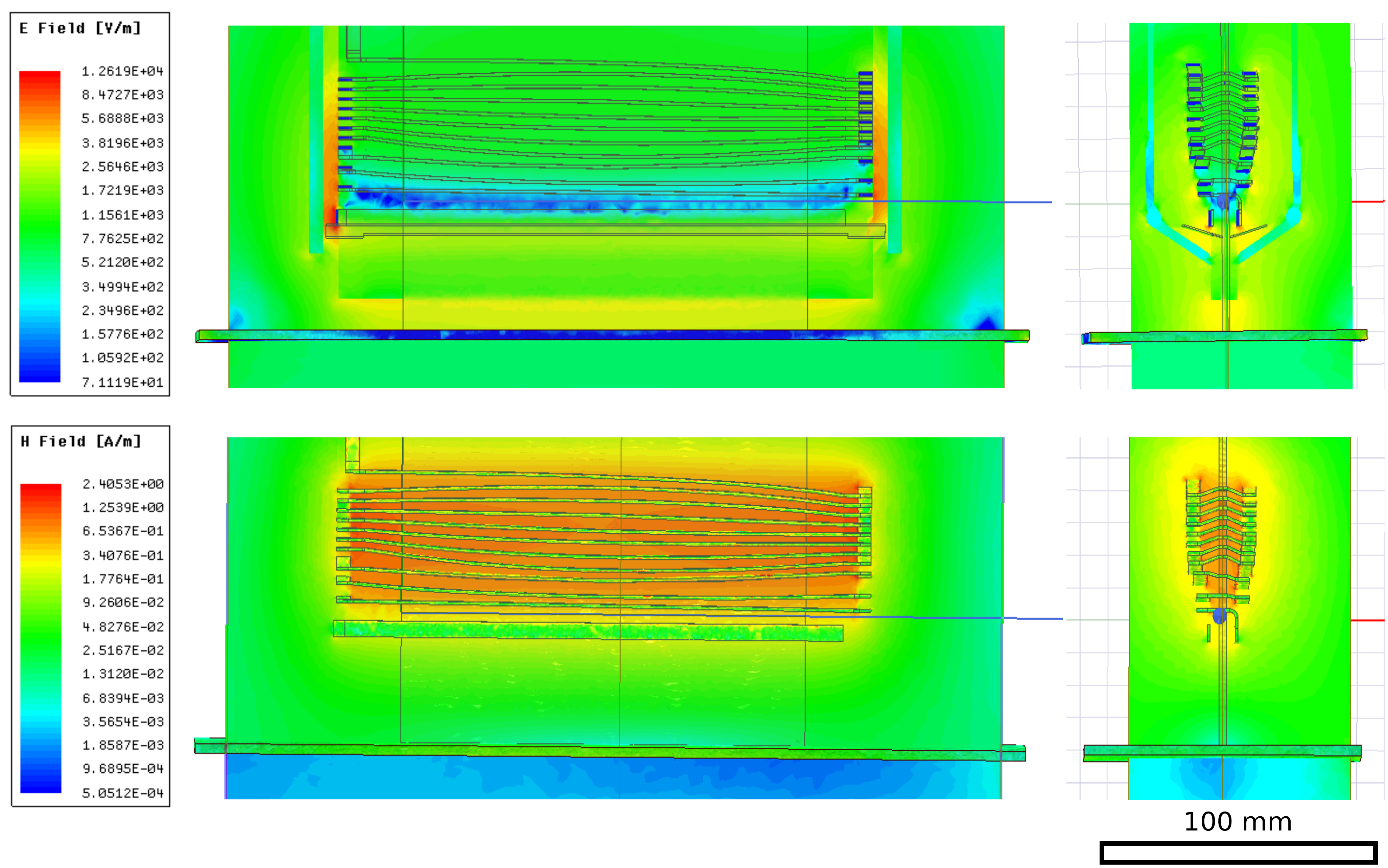}
 \centering
 \caption{Front and side views of the electromagnetic field components ($E$ in top, $H$ in bottom) of the RF plasma slit jet before plasma ignition. It was obtained by the numerical simulations for power 600\,W and frequency 13.56\,MHz. The distance between the slit exit and grounded Al plate covered with 3 mm thick dielectric substrate ($\epsilon_\textrm{r}$ = 6) was 10 mm.}
 \label{fig:elmagfield}
\end{figure}

A simple numerical model of the spatial distribution of the electric and magnetic field of the PSJ positioned above a grounded Al plate covered with dielectric substrate  (figure~\ref{fig:elmagfield}) has revealed a distribution of electromagnetic (EM) field with both the components having non-negligible intensity ($E_\mathrm{max} = 1.3\cdot10^{4}$\,V/m, $H_\mathrm{max} = 2.4$\,A/m). The simulation of the PSJ with plasma approximated as a conductor in a quasi-stationary state yields similar values (see Supplementary Information, section~S1). Higher electric field intensity is induced in regions between the winding of high-voltage RF electrodes and the ground (either the PSJ metal cover or the substrate). The intensity pattern of the $E$ formed on the slit sides directly below the last turn of the high-voltage RF electrode winding is similar to the one reported by Bourdon~\textit{et al.}~\cite{Bourdon2016} for the capacitively coupled plasma ignited inside a quartz capillary with a high-voltage outer ring electrode. Therefore, the PSJ is likely a capacitively coupled discharge. This conclusion is further confirmed by the comparison of the PSJ simulated $E$ values ($E_\mathrm{max} = 1.3\cdot10^{4}$\,V/m) with the electric field intensity range $E = 10^4-10^6$\,V/m reported for various capacitively coupled atmospheric plasma jets~\cite{Janca_SCT99,liu2021,Bourdon2016}. The simulated PSJ $E$ is at the lower end of the $E$ values in the literature because it only accounted for the externally applied electric field. The reported  values, measured or simulated for the active discharges, were the sum of the externally applied electric field and the electric field induced by the charge particles generated inside the plasma column. 

According to the simulation, the magnetic field $H$ was induced inside the PSJ body, primarily in the core area of the resonance coil. The maximum $H$ of the PSJ, 2.4\,A/m, is three to four orders of magnitude lower than in RF inductively coupled plasmas $H = 10^3-10^4$\,A/m~\cite{boulos1976,xue2001}, yet it is not zero as in the purely capacitively coupled discharges, suggesting a possibility of the PSJ discharge having a small inductive component. 

Two regions are crucial for the proper function of the PSJ: (i) the higher electric field intensity region between the winding of the high-voltage RF electrode and the PSJ metal cover and (ii) the lower intensity region formed above the last turn of the high-voltage RF electrode. The relatively high $E$ in region (i) reveals a potential weak point in the construction. The calculated electric field is not high enough to induce the breakdown of ambient air filling the cavity surrounding the discharge slit. However, an unwanted arc ignition can occur if the resonance matching plate edge comes too close to the coil winding or if a poorly structured surface reflects Ar flow to the region (i). Therefore, we continuously purge the cavity with nitrogen during the PSJ operation to limit the possibility of breakdown outside the discharge slit. The region (ii) with the relatively low electric field intensity was created intentionally, through coil turns orientation and distance variation, to prevent plasma propagation deep into the slit body.

%%Obrázek
%%Obrázek

\subsection{Self-organization patterns}
\begin{figure}[b]
\includegraphics[width=\textwidth]{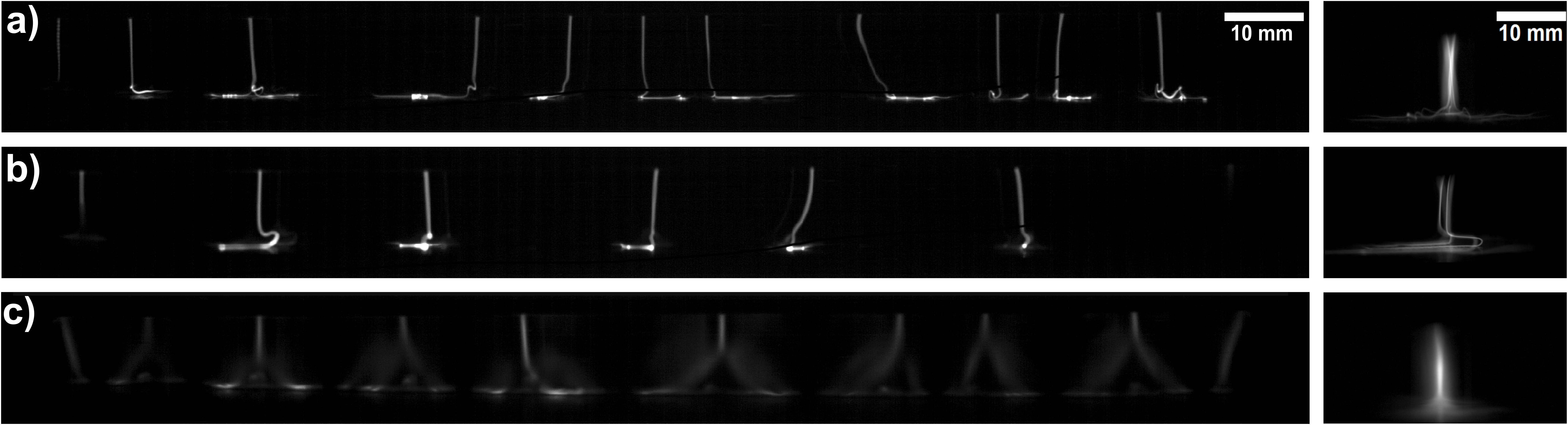}
\centering
\caption{Front (right) and side (left) views of the RF plasma slit jet discharge ignited in three different gas mixtures at 500\,W: (a) 67 slm Ar, (b) 67\,slm Ar/1\,slm O$_2$, and (c) 67\,slm Ar/1.5\,slm N$_2$. The distance between the slit outlet and the dielectric surface (mica composite, thickness of 6\,mm) was kept at 10\,mm.}
\label{fig:discharge}
\end{figure}

\begin{figure}[p]
\includegraphics[width=\textwidth]{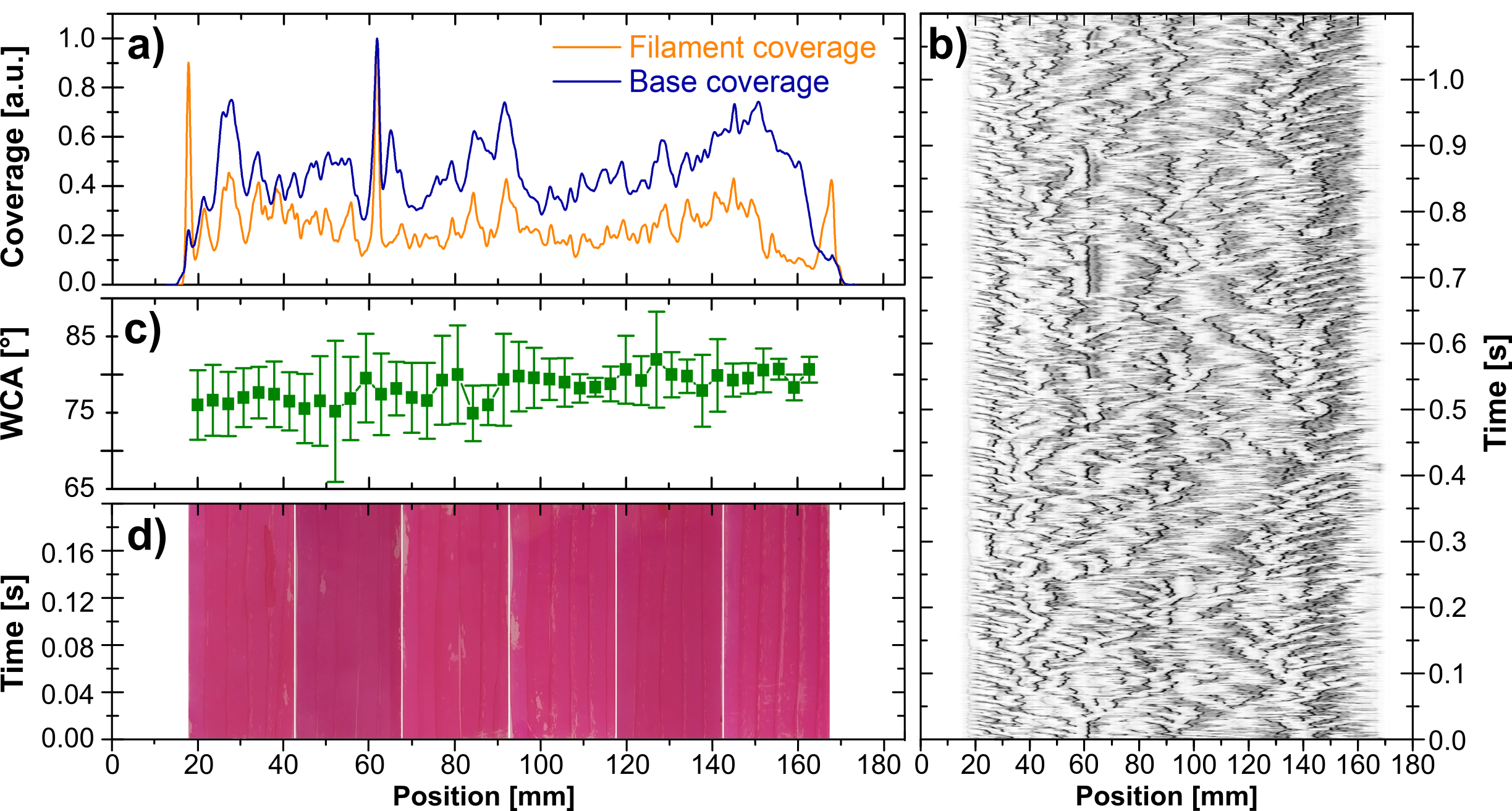}
\centering 
\caption{Evaluation of the filament behavior in the Ar PSJ discharge (67\,slm Ar, 500\,W) moving at 100\,mm/s above the mica composite: (a) filament entropy distribution calculated for the interval of 1.1\,s at the slit exit (filament coverage) and contact with the mica (base coverage); (b) the base coverage plotted over time in the inverted color scheme (plasma filaments are dark). The uniformity of the PP treatment determined for the faster PSJ speed, 250\,mm/s, above the PP: (c) WCA dependence on the position along the slit; (d) treatment uniformity visualized by the ink test, Quicktest 38.}
\label{fig:Ar-homogeneity}
\end{figure}

\begin{figure}[p]
\centering
\includegraphics[width=0.98\linewidth]{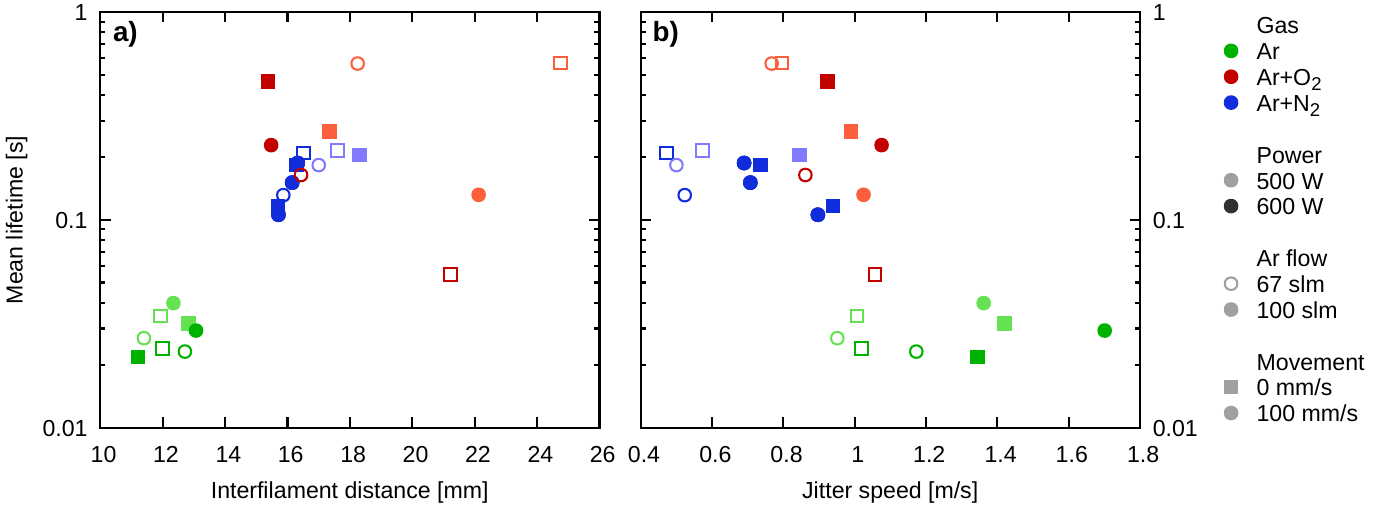}
\caption[width=0.9\textwidth]{Correlation between the lifetime and (a) the inter-filament distance or (b) the average travel speed (jitter speed) of filaments along the length of the slit for different gas mixtures, RF power, and Ar flow rates when the PSJ was stationary or moving with 100\,mm/s relative to the dielectric plate beneath the jet.}
\label{fig:filament-summary}
\end{figure}

For all the studied working gas compositions (Ar, Ar/N$_2$, and Ar/O$_2$), the PSJ discharge maintained a certain degree of filamentary character with stationary filaments situated at both ends of the slit. The pure Ar PSJ filaments (the front view in figure~\ref{fig:discharge}a) were the thinnest (tenths of mm) and the most numerous, 12--15 per the slit length, with the mean inter-filament distance of 12\,mm (figures~\ref{fig:Ar-homogeneity}a and~\ref{fig:filament-summary}a). In contact with a dielectric substrate, the filaments created a 20--30\,mm wide base spread equally to both sides (side view in figure~\ref{fig:discharge}a). 
%predelano - KATKA
The Ar PSJ filaments changed positions over time, shuffling along the length of the slit at the jitter speed of 0.9--1.7\,m/s %jitter speed values in the figure 6b do not correspond with results shown for each analyzed condition in the report_v2 (there, it is 0.8--1.8 mm/s)
(figures~\ref{fig:Ar-homogeneity}b and \ref{fig:filament-summary}b). The variance in the jitter speed was primarily caused by the varied Ar flow rates. The filaments moved faster at the higher Ar flow rate of 100\,slm. In addition to the motion, filament formation and annihilation events occurred. 

The self-organization of filaments into patterns defined by a characteristic inter-filament distance resembles the self-organizing patterns of a quasi-1D DBD described and explained by Boeuf~\textit{et al.}~\cite{Boeuf2012}. Their self-contained fluid discharge model showed that each DBD filament was surrounded by an inhibition region formed by charge spreading along the dielectric surfaces. In this region, the ignition of another filament was not allowed, and low-current (Townsend) side discharges were formed on its side due to the combination of the transverse plasma diffusion and ion-induced secondary emission. The side discharges were essential for filament generation, annihilation, and pattern formation. 

%Obrázky
%Obrázky

%Konec obrázků
%Konec obrázků

In the quasi-1D DBD system with boundaries, the existence of inhibition zones causes filaments to rearrange following the spatial period intrinsic to the activator-inhibitor system, \textit{i.\,e.}, filaments maintain a characteristic inter-filament distance. We observe the same self-organizing behavior in our RF PSJ discharge and, accordingly, attribute it to the filaments' ion sheaths spreading the charges along the dielectric surface of the slit. The formation of low-current side discharges is likely vital to both the generation and the motion of PSJ filaments. Low current side discharges could evolve into plasma filaments if the applied voltage is sufficiently large. As the Ar PSJ plasma is easy to sustain (applied voltage is much higher than the breakdown), this transition/new full filament formation happens frequently. However, such newly created filaments disrupt the characteristic inter-filament distance, forcing the system to rearrange. 
%DODELAT Katko, skutecne merge? To jaksi odporuje tomu, ze by mohly dva filamenty prijit blizko k sobe. Lenka: preformulovala jsem to
If filaments come too close, one of them usually decays. The frequent ignition and decay processes led to short filament lifetimes (0.020--0.035\,s), a behavior specific to the Ar PSJ discharge (figure~\ref{fig:filament-summary}a). The observed filament motion is possibly a result of the asymmetry in low current side discharges induced by the generation and annihilation of the fully formed filaments.

%Obrázek
%Obrázek

 \begin{figure}[b]
 \includegraphics[width=\textwidth]{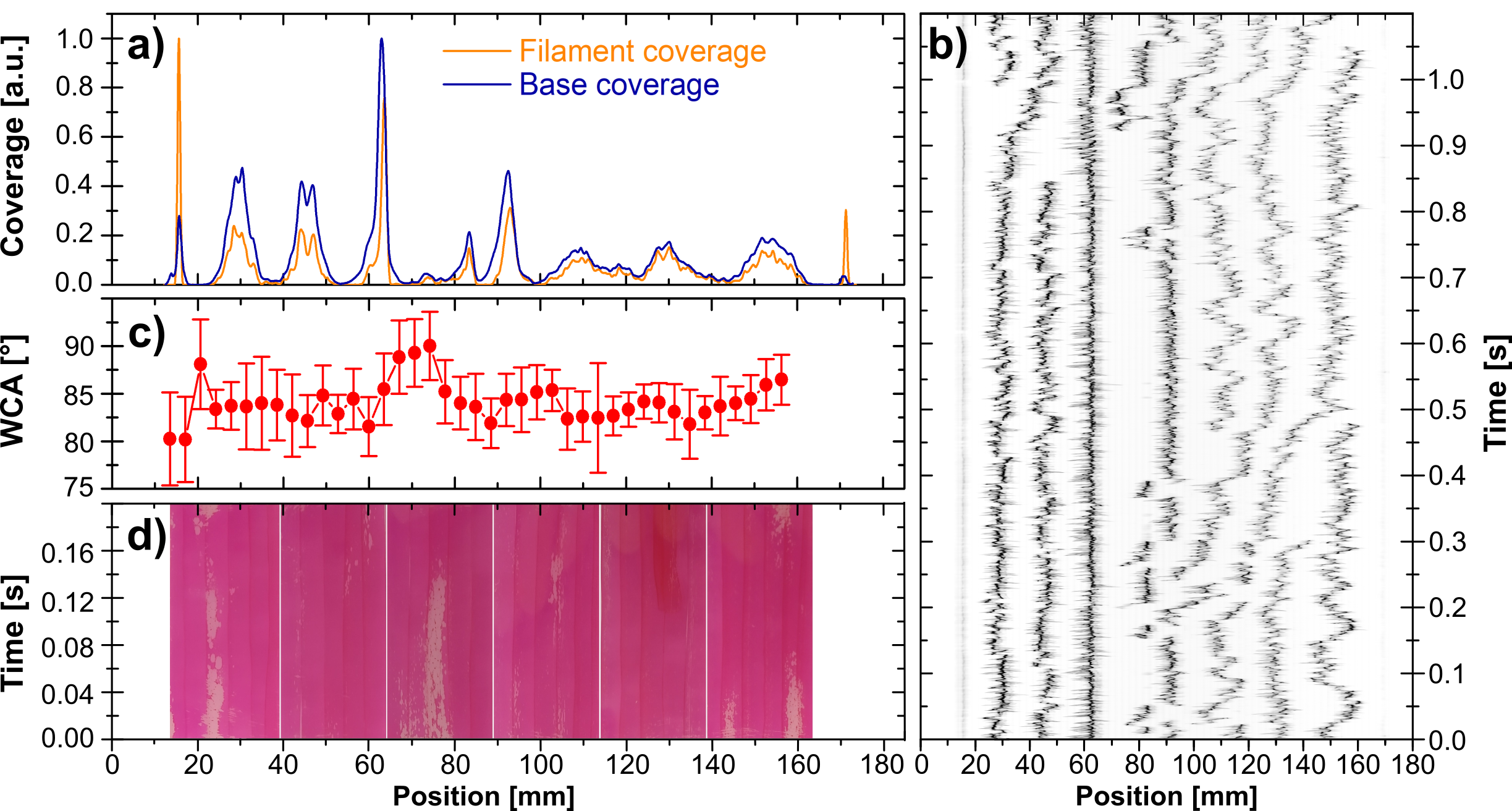}
 \centering
 \caption{Filament characteristics of Ar/O$_2$ PSJ discharge (67\,slm Ar/1\,slm O$_2$, 500\,W) moving at 100\,mm/s above the mica composite: (a) filament entropy distribution calculated for the interval of 1.1\,s at the slit exit (filament coverage) and contact with the mica (base coverage); (b) the base coverage plotted over time in the inverted color scheme (plasma filaments are dark). The uniformity of the PP treatment determined for the faster PSJ speed, 250\,mm/s, above the PP: (c) WCA dependence on the position along the slit; (d) treatment uniformity visualized by the ink test, Quicktest 38.}
 \label{fig:Ar_O2-homogeneity}
 \end{figure}

%Konec obrázku
%Konec obrázku

The discharge ignited in the Ar/O$_2$ mixture was similar in appearance to pure Ar (figure~\ref{fig:discharge}b) with thin, well-defined filaments. In contact with a dielectric substrate, all the filaments in Ar/O$_2$ had a specific S-like shape with a preferential orientation. The S-like shape appeared for all the studied conditions (500 or 600 W, Ar flow rate 67 or 100 slm) if oxygen was added (1 slm).
%result of gas dynamics. %Dokážeme vysvětlit proč to tak je? %Podívat se, jestli účka vznikají i při interakci s polypropylenem? => Ano. 
The number of filaments was lower in the Ar/O$_2$ mixture than in pure Ar discharge, 7--11 per the slit length. Accordingly, the mean distance of the closest filaments increased to 15--26\,mm (figures~\ref{fig:filament-summary}a and \ref{fig:Ar_O2-homogeneity}a). The decrease in the filament number is connected with the electronegative and molecular character of O$_2$ plasma. 
%DODELAT Katko, myslim, ze krome elektronegativniho charakteru O2 by se nemelo zapomenout na to, ze je to molekularni plyn, podobne jako N2 - disociace a excitace. Nemyslis? Text jsem prepsala Katka: OK, souhlasim 
Electrons sustaining the discharge are lost in the attachment processes and their energy is lost by dissociation and excitation of molecules. Therefore, the electron density in Ar/O$_2$ discharge should be lower than in pure Ar~\cite{Park_2008}. Decreasing the number of electrons while keeping or lowering their energy reduces the number of ionization collisions they initiate, thus, limiting the number of filaments sustained by the given applied power. 
%This correlation is nicely illustrated by changing the power (i.\,e. electron density) as the number of filaments was higher at 600 W. 
Unlike the fast-traveling short-lived filaments of the Ar PSJ, the filaments of the Ar/O$_2$ discharge were almost stationary (figure~\ref{fig:Ar_O2-homogeneity}b) with long lifetimes (figure~\ref{fig:filament-summary}). The contrasting behavior of the Ar and Ar/O$_2$ PSJ filaments results from differences in plasma sustainabilities. At the studied powers, the probability of low-current side discharges transitioning into the full plasma filaments is limited in the Ar/O$_2$ discharge. Consequently, the self-organized pattern is seldom disrupted, leading to lesser movement and longer lifetimes.

 \begin{figure}[b]
 \includegraphics[width=\textwidth]{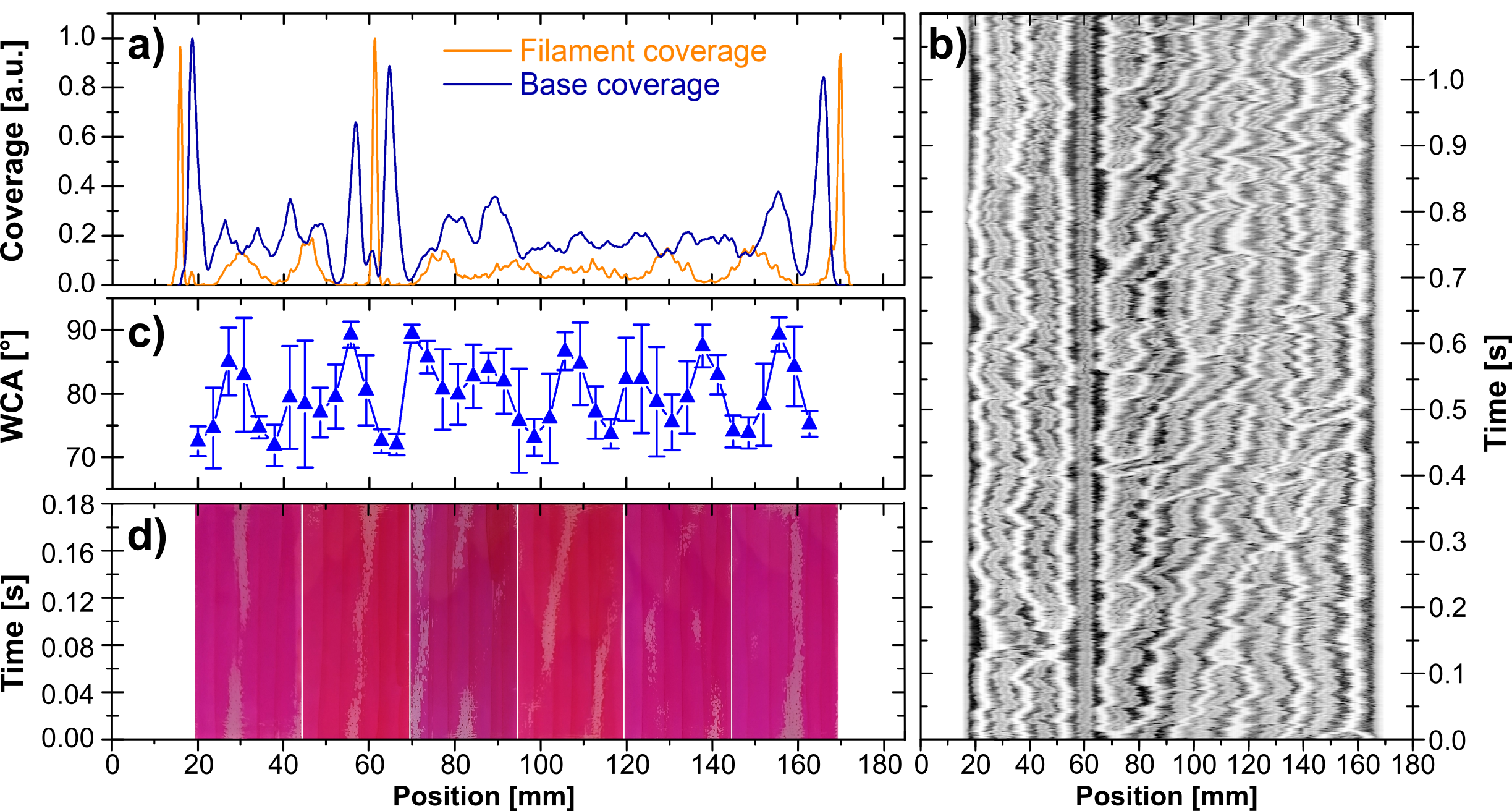}
 \centering
 \caption{Filament characteristics of Ar/N$_2$ PSJ discharge (67\,slm Ar/1.5\,slm N$_2$, 500\,W) moving at 100\,mm/s above the mica composite: (a) filament entropy distribution calculated for the interval of 1.1\,s at the slit exit (filament coverage) and contact with the mica (base coverage); (b) the base coverage plotted over time in the inverted color scheme (plasma filaments are dark). The uniformity of the PP treatment determined for the faster PSJ speed, 250\,mm/s, above the PP: (c) WCA dependence on the position along the slit; (d) treatment uniformity visualized by the ink test, Quicktest 38.}
  \label{fig:Ar_N2-homogeneity}
 \end{figure}
 
In the Ar/N2 mixture, the constricted plasma filaments, typical of other gas feeds, became shorter (figure~\ref{fig:discharge}c) with the increasing N$_2$:Ar ratio until they sustained only inside the slit. Simultaneously, a peculiarly-shaped diffuse plasma region developed. Its color was akin to the nitrogen orange afterglow~\cite{piper1994, bravo2015}. Contrary to the expectations based on the appearance of Ar/N$_2$ capillary discharges with visible nitrogen afterglow~\cite{bravo2015, munoz2019, yehia2020}, the diffuse plasma was not present directly below the constricted filament but on its sides, creating a double plume structure (total diameter of 15--20\,mm). 

%DODELAT ten puvodni popis se mne Katko nezda, ze by plne popisoval realitu: 
% difuse plasma is not DIRECTLY below plasma filament
% ackoliv nekdy je filament na jedne strane trojuhelniku silnejsi, popis mne nedaval smysl z hlediska tvaru
% dalsi odstavec zminuje, pokud jsem to dobre pochopila, ze se mohlo stat, ze filament normalne dosahl na substrat. Do SI by to chtelo pridat vice obrazku.
%Katka: Do SI jsem pridala obrazky, na kterých se jasne da odlisit difuzni afterglow pod diskretnim filamentem a difuzni plasma na jeho stranach, coz jsem se snazila evidentne neuspesne popsat v testu
%Contrary to the expectations based on the appearance of Ar/N$_2$ capillary discharges with visible nitrogen afterglow~\cite{bravo2015, munoz2019, yehia2020}, the diffuse plasma was not only present directly below the constricted filament but also on its sides, creating a double plume-like shape (total diameter of 15--20\,mm). Unexpectedly, the diffuse extension of discrete filament quickly randomly merged with one of the diffuse side plumes, leaving a dark triangular region in the middle (Figure~\ref{fig:discharge}c).

%%%%A dark space has also formed on the interface of the diffuse envelope and the constricted filament. Generally, the transition of the constricted filament to the diffuse filament to the part of the diffuse plume was smooth without any darker spaces. 

The structure of the Ar/N$_2$ plasma in contact with the mica dielectric substrate depended on the length of the constricted filaments. When they were long enough to maintain contact with the substrate, the plasma was formed by constricted filaments, but unlike in Ar or Ar/O$_2$, they were surrounded by diffuse plasma plumes.
%DODELAT Katko, toto je tvuj puvodni text: the plasma had a relatively common appearance of the constricted filaments surrounded by diffuse plasma envelopes. Zda se mne, ze by se melo zduraznit, ze v Ar/N2 jako jedinem plynu byly ty difuzni casti. Je to tak, ze? Taky by bylo fajn pridat do SI obrazek, jak to vypada, kdyz jsou ty Ar.N2 filamenty kratke. Katka: Ano, jedine u Ar/N2 vyboje vidime ty difuzni casti. Do SI pridan obrazek ilustrujici, jak se se zvysujicim průtokem N2 meni tvar diskretni a difuzni casti filamentu v kontaktu s podlozkou.
In the case of shorter filaments, their diffuse extension merged with the diffuse side plumes, leaving a dark triangular region in the middle (figure~\ref{fig:discharge}c). A plasma vortex could form in the middle dark triangular region. Its existence and rotational direction depended on the PSJ settings (Ar flow rate, power, distance between the slit exit and the substrate). From the side view (right column in figure~\ref{fig:discharge}c), the Ar/N$_2$ diffuse plasma formed a symmetrical base in contact with the mica substrate, its diameter being the smallest of all the studied gas mixtures (maximum of 10\,mm).

Following Boeuf \textit{et al.}~\cite{Boeuf2012}, we can describe the appearance of the Ar/N$_2$ discharge as a central plasma filament surrounded by an inhibition region with low current (Townsend) side discharges on the sides. It is consistent with the Ar and Ar/O$_2$ PSJ discharges, as the same structure was used to described their self-organized filament patterns, the only difference being the visibility of the low-current side discharges. In the Ar and Ar/O$_2$ plasma, the side discharges were invisible, whereas in the Ar/N$_2$ PSJ they had a form of a red-to-orange colored diffuse plasma plumes. 

Given the similarities with the nitrogen orange afterglow~\cite{piper1994, bravo2015}, the formation of Ar/N$_2$ specific diffuse side plumes is likely connected to the presence of long-lived nitrogen species, such as ground-state atoms N($^4$S) or atomic (N($^2$D), N($^2$P)) and molecular (N$_2$(A), N$_2$(B), N$_2$(C)) metastable states. As the named species are neutral, their diffusion is not limited by the electric field, allowing them to transverse the inhibition region freely. Beyond the inhibition zone, they can help to sustain the diffuse Townsend side discharges by inducing secondary emission of electrons from the dielectric surface of the slit. Such electrons have low energies and, therefore, are incapable of many ionizing collisions, keeping the charge density of side discharges low and limiting their development into full filaments. This process is analogous to the working principle of the homogeneous DBD in N$_2$~\cite{massines2012, Massines2009}. There, the memory effect (\textit{i.\,e.}, production of electrons between the pulses) necessary for a homogeneous DBD formation is driven by secondary emission induced by N$_2$(A) metastable impacts. 

%DODELAT Ja bych tenhle text uplne vynechala. Jak pise David, divne se to prekryva s jiz publikovanou literaturou. Nahore jsem smazala ten diffuse filament, pro me divny pojem, tady jeste navic diffuse full filament ... Tenhle text zbytecne vse komplikuje a neprinasi zadne rozumne vysvetleni. Katka: dobre, muzeme
%A fraction of the long-lived nitrogen species created inside the discrete PSJ filament region and its vicinity followed the direction of the gas flow, forming a post-discharge (afterglow) diffuse plume that randomly merged with the diffuse side plumes. The plasma channel fusion increased the density of charged particles, causing the plasma column to constrict partially. The merging is not permanent as the diffuse full filament head changes the orientation at random time intervals, probably due to some type of discharge instability. %(flow, thermal, electronic, ... ).   

The number of Ar/N$_2$ PSJ filaments, 8--9, remained constant regardless of the applied power and the gas feed flow rates. Accordingly, the range of average distances between the discrete filaments was narrow, 16 to 18\,mm (figure~\ref{fig:filament-summary}a). The fixed spatial period of the Ar/N$_2$ filament system could result from charges deposited on the dielectric surface by the wide low-current side discharges. Additionally, the studied applied powers (500 and 600\,W) could be too low to generate more than the observed 8--9 filaments, as a large part of the supplied energy was likely used to excite numerous rotational and vibrational levels of N$_2$ molecules rather than generate charged particles. A low charge density limits the transition of side discharges into a full filament, resulting in a stable pattern with long filament mean lifetimes, 0.25 -- 0.1\,s, and low jitter speed, 0.4--1\,m/s (figure~\ref{fig:filament-summary}b).

%Obrázek Souhrn pro vsechny smesi

\begin{figure}[t]
\centering
\includegraphics[width=0.98\linewidth]{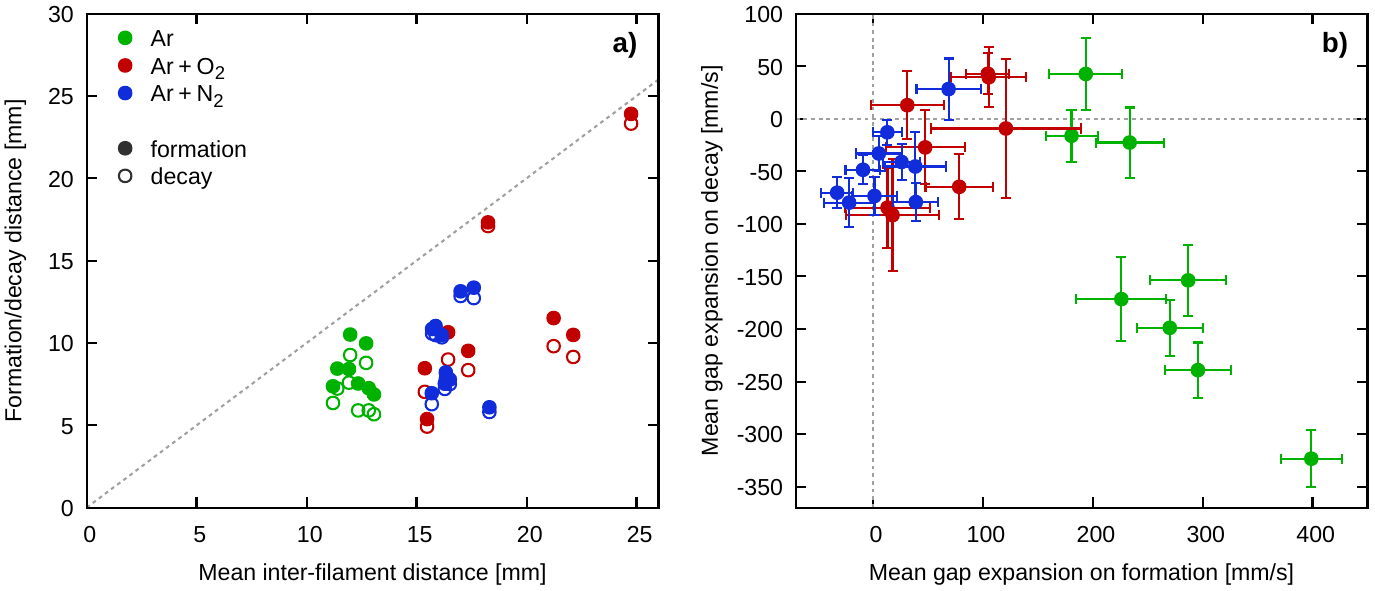}
\caption[width=0.9\textwidth]{Inter-filament distance upon formation and decay: (a) comparison of the mean distance upon formation and decay with the overall mean inter-filament distance, the dashed line corresponding to the distances being equal; (b) evolution of the distance between filament's nearest neighbors (gap) during formation and decay.}
\label{fig:formation-decay}
\end{figure}

%Konec obrázku

%Souhrn pro vsechny smesi
The overview of self-organization characterized by an inter-filament distance in all studied gas feeds (Ar, Ar/O$_2$, and Ar/N$_2$) is illustrated in detail in figure~\ref{fig:formation-decay}. Figure~\ref{fig:formation-decay}a gives a static picture, comparing the inter-filament distance upon formation/decay to the mean distance. Both formation and decay distances are smaller than the mean distance, in some cases considerably. The decay distances seem overall slightly smaller than formation distances. Figure~\ref{fig:formation-decay}b gives a complementary dynamic picture, showing how the gap between filament's nearest neighbors is expanding (positive sign) or contracting (negative sign) during filament formation and decay. If the gap expands during formation and contracts during decay, the data should lie in the fourth quadrant of the plot. It is only mostly true since in some cases the gap expansion rates have small values of the opposite sign than expected.  Nevertheless, the expansion rates in Figure~\ref{fig:formation-decay}b have significant uncertainties, and all data lie inside or close to the fourth quadrant (uncertainties in figure~\ref{fig:formation-decay}a are too small to show). Figure~\ref{fig:formation-decay}b thus still supports the same conclusions on self-arrangements of filaments demonstrated above as the minimum distance in most inter-filament distance distributions.

\subsection{Plasma treatment uniformity}
\label{sec:homo}
\begin{figure}[b]
 \includegraphics[width=\textwidth]{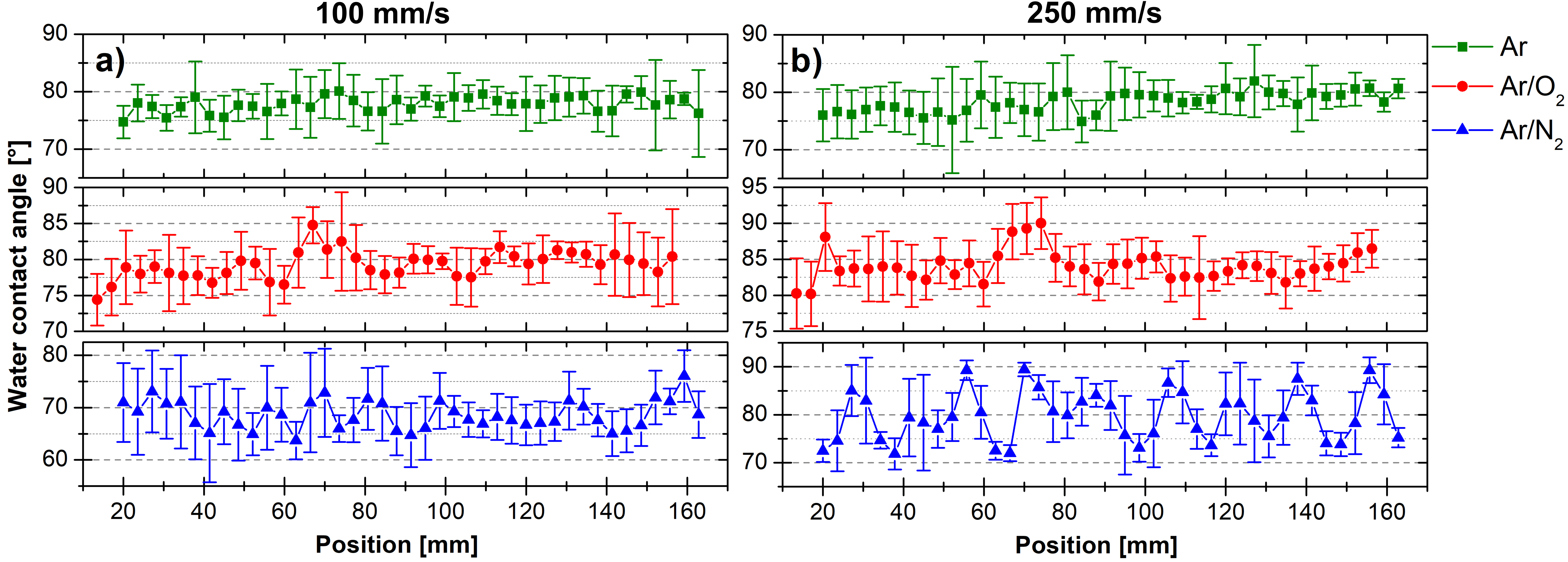}
 \centering
 \caption{Dependence of water contact angles on the position along the jet slit for (a) 100\,mm/s and (b) 250\,mm/s treatment speed. A slight shift in the position of the points measured for Ar/O$_2$ working gas mixture was due to the samples misalignment, the total length of the samples was 150\,mm to the length of the discharge $\sim$160\,mm.}
 \label{fig:WCA}
 \end{figure}

The filamentary character of the discharge raises the question of plasma treatment uniformity, which will be discussed in the example of polypropylene.
At the lower movement speed (100\,mm/s), the 38\,mN/m ink test did not reveal any insufficiently treated areas along the entire length of the slit for any of the tested feed gases. The water contact angle (WCA) measurement offers more precise information, as the results can show the whole scale of the surface free energy (SFE) values and not just the visualization of the part of the surface with the SFE values lower or higher than a certain number. For the PP treated at 100\,mm/s in the Ar/O$_2$ or Ar/N$_2$ mixtures, the WCA results (figure~\ref{fig:WCA}a) showed slight inhomogeneities that became more pronounced at the higher movement speed (figure~\ref{fig:WCA}b). At the 250\,mm/s, the SFE of these areas decreased below the 38\,mN/m, and their time-dependent evolution across the slit could be visualized by the ink test results (figure~\ref{fig:Ar_O2-homogeneity}d,~\ref{fig:Ar_N2-homogeneity}d).

%Obrázek
%Obrázek

  %nedát tento obrázek do přílohy?
 
%Konec obrázku
%Konec obrázku

The plasma treatment of PP in the Ar PSJ discharge was the most uniform one (figures~\ref{fig:Ar-homogeneity}c,d and \ref{fig:WCA}), despite the discharge's filamentary appearance. The analysis of fast camera images proved to be invaluable for the explanation of these unexpected results. The time-dependent base coverage plot (figure~\ref{fig:Ar-homogeneity}b) shows that the high treatment uniformity is a result of the short filament lifetimes. Quick formation and decay of filaments led to a higher degree of randomness in the positions along the slit and faster jitter speeds caused by the self-organizing nature of filaments trying to maintain an ideal inter-filament distance. The overall base coverage plot (figure~\ref{fig:Ar-homogeneity}a) is a fast camera analysis analogy to the surface analysis results. For the Ar discharge, it confirmed the excellent filament distribution uniformity over the whole slit width. The narrow jutting peak at $\sim$60\,mm is a result of recurring longer-lived stationary filaments. They are fixed at the position by a higher electromagnetic field, induced by a slight bump in the otherwise regularly distanced slit. 

The plasma treatment in the Ar/O$_2$ mixture was relatively uniform with a few less treated areas, namely between 65 and 80\,mm (figure~\ref{fig:WCA}). It resulted from the already mentioned bump inside the slit cavity that fixed a stationary plasma filament at the $\sim$60\,mm position. Due to the longevity of Ar/O$_2$ filaments, anchoring one in a place had a more substantial influence on the overall discharge appearance than in the case of shorter-lived Ar filaments. The stationary filament at the narrowed slit region served as a barrier to the movement of neighboring filaments (figure~~\ref{fig:Ar_O2-homogeneity}b). Their positions also became fixed, as the self-organizing nature of the discharge kept them at the ideal interval from the central filament, and a formation of new filaments seldom led to a disruption of the inter-filament distance around the narrowed slit region. 
The less treated area appeared only on the side of the fixed filament at 60\,mm that was further away from the slit end, \textit{i.\,e.}, at 65--80\,mm (figure~\ref{fig:Ar_O2-homogeneity}d). It probably stems from a combination of the bump shape and the effect of the fixed filament at the slit exit, squishing the other filaments in the region between 20 and 60\,mm together. Likewise, the anchoring of filaments to the slit ends lead to a regular appearance of less treated areas at 25 and 160\,mm.

%DODELAT Katko, nemelo by se tady ale zminit, ze na okraji 25mm a 160 mm se taky pravidelne objevuje mene modifikovany povrch? Katka: pridana veta na konec predchoziho odstavce
The treatment nonuniformities that sporadically appeared between 80 and 160\,mm (figure~\ref{fig:Ar_O2-homogeneity}d) are created during a filament number change. Random decay or ignition events can lead to the formation of brief windows with greater inter-filament distances since it takes some time before self-organization is established again.  Interestingly, the preferred filament positions are well defined in the overall base coverage plot (figure~\ref{fig:Ar_O2-homogeneity}a) with a nearly zero coverage in between the intensity peaks. The results are in strict contrast to treated surface analyses that yielded a much more uniform picture (figure~\ref{fig:Ar_O2-homogeneity}cd). The discrepancy can be explained by an insufficient dynamic range of the fast camera images, making the low-intensity plasma in between the filaments invisible at the used camera settings. Alternatively, similar to diffuse plasma of Ar/N$_2$ discharge, the surface treatment of the in-between areas could have been induced by a cloud of non-radiative neutral reactive species (Ar metastables, OH ground-state molecules, \textit{etc.}).

% ja bych ty difuzni casti plazmatu nenazyvala diffuse filaments
The Ar/N$_2$ PSJ with the broad diffuse plumes was visually the most homogeneous discharge. However, the surface treatment was the least uniform of all the tested gas feeds, with less-treated areas (2--5\,mm in diameter) present along the whole length of the slit (figure~\ref{fig:Ar_N2-homogeneity}c,d). A comparison of the Quicktest 28 ink test result (figure~\ref{fig:Ar_N2-homogeneity}d) with the base coverage map (figure~\ref{fig:Ar_N2-homogeneity}b) correlated the presence of these areas to dark regions separating the individual filaments.  In contrast to the surface treatment analysis results, the overall base coverage (figure~\ref{fig:Ar_N2-homogeneity}a) showed only three regions with lower intensity in the positions next to the permanently fixed filaments. The presented false uniformity of the overall coverage is caused by the constant movement of the remaining filaments (traveling along the slit and oscillating in place). The WCA results were not so impacted (high error bars of certain values in figure~\ref{fig:Ar_N2-homogeneity}c) as the treatment time (0.25\,s) was short compared to the fast camera recording interval (1.09\,s). Comparing the gas feeds, the Ar/N$_2$ (figure~\ref{fig:Ar_N2-homogeneity}b) PSJ shared the most features with the Ar/O$_2$ (figure~\ref{fig:Ar_O2-homogeneity}b) discharge, illustrating that the main factor influencing the filament behavior and surface treatment uniformity is their lifetime (ignoring the influence of the slit cavity defect at $\sim$60\,mm).

\begin{figure}[tbp]
 \includegraphics[width=0.7\textwidth]{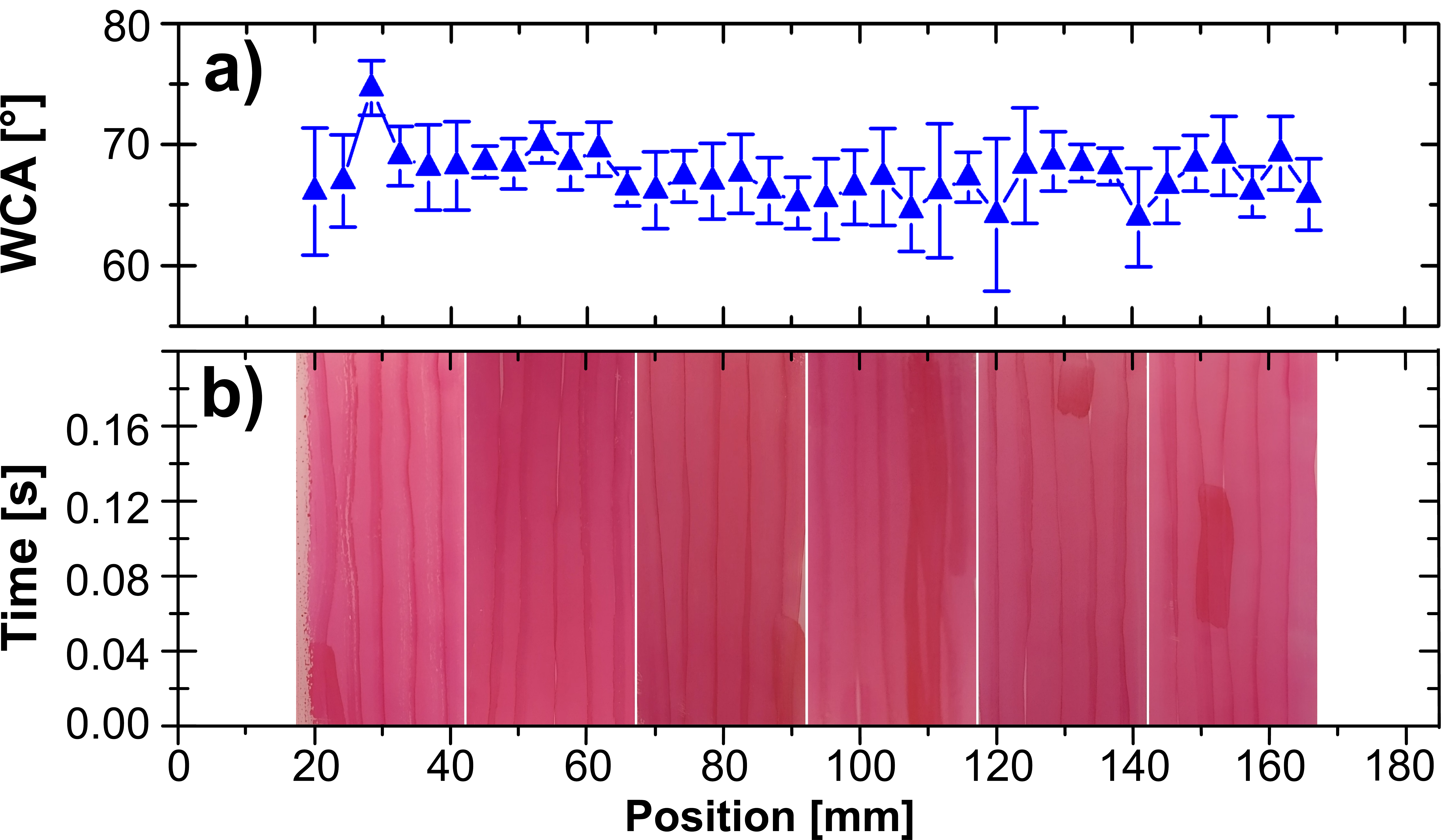}
 \centering
 \caption{Ar/N$_2$ discharge (100\,slm Ar/3\,slm N$_2$, 600\,W, 100\,mm/s -- 2x). (a) Dependence of the WCA on the position along the slit. (b) The uniformity of the treatment visualized by the ink test Quicktest 38.}
 \label{fig:100mms-2x-ArN2-homogeneity}
 \end{figure}

The above-discussed results on plasma filament separation and treatment uniformity in Ar/O$_2$ and Ar/N$_2$ might look contradictory. Even though the Ar/O$_2$ filaments (figure~\ref{fig:Ar_O2-homogeneity}) are separated by much larger dark region than in the Ar/N$_2$ discharge (figure~\ref{fig:Ar_O2-homogeneity}), the plasma modification uniformity was better using the Ar/O$_2$ PSJ. The difference can be explained by a difference in the reactive gas compositions. The Ar/O$_2$ plasma produces a higher amount of reactive oxygen species (such as O and OH radicals), whereas reactive nitrogen species dominate the gas chemistry of the Ar/N$_2$ PSJ. The PP reactivity with nitrogen gas species is much lower than with the oxygen ones~\cite{dorai2003}, and the nitrogen functional groups are less hydrophilic than oxygen ones. Thus, a much higher number of reactive nitrogen gas species is needed to induce the same level of increase in the surface free energy of PP. The nonuniformities observed on the PP treated in the Ar/N$_2$ discharge were removed by increasing the number of passes to two (figure~\ref{fig:100mms-2x-ArN2-homogeneity}). In these conditions, the PSJ was successfully used to increase PP adhesion to the epoxy glue, and Ar/N$_2$ plasma was the most efficient one~\cite{polaskova2021}.

%Obrázek
%Obrázek

%Obrázek
%Obrázek

\section{Conclusion}
\label{sec:conclusion}
The RF plasma slit jet (PSJ) produces 150\,mm wide streaming plasma outside the jet body. The coupling is achieved through an RF coil designed around the dielectric slit inserted in a tunable metal cavity. The simulation of the electromagnetic field without plasma showed that a high electric field is induced in the regions between the winding of high-voltage RF electrodes and the ground (either the PSJ metal cover or the substrate). A non-zero magnetic field suggests that, besides the primary capacitive coupling, there is a small inductive component. For all the studied gas feeds (Ar, Ar/N$_2$, and Ar/O$_2$), the PSJ discharge maintained a certain degree of filamentary character with stationary filaments situated at both ends of the slit. The self-organization of filaments into patterns defined by a characteristic inter-filament distance resembles the self-organized patterns of dielectric barrier discharge (DBD) filaments. Similarly, as in DBD, the filaments are surrounded by an inhibition zone that does not allow two filaments to come closer to each other.

The pure Ar PSJ filaments were the thinnest (tenths of mm) and the most numerous. As the breakdown voltage in pure Ar is relatively low compared to the applied one, new filaments formed frequently. Such newly created filaments disrupted the characteristic inter-filament distance, forcing the system to rearrange. 
%DODELAT i zde je to merge, nejak rozresit - Lenka: preformulovano
If filaments come too close, one of them usually decays. The frequent ignition and decay processes led to short filament lifetimes (0.020--0.035\,s) and their high jitter speed (0.9--1.7\,m/s), a behavior specific to the Ar PSJ discharge. The number of filaments was lower in the Ar/O$_2$ and Ar/N$_2$ mixtures. It was attributed to a loss of energy in dissociation and excitation of numerous rotational and vibrational levels and a loss of electrons due to oxygen electronegativity. Since the probability of low-current side discharges transitioning into the full plasma filaments was limited in the gas mixtures, the self-organized pattern was seldom disrupted, leading to lesser movement and longer lifetimes.

The structure of the Ar/N$_2$ plasma in contact with the mica dielectric substrate depended on the length of the constricted filaments. When they were long enough to maintain contact with the substrate, the plasma was formed by constricted filaments, but unlike in Ar or Ar/O$_2$, they were surrounded by diffuse plasma plumes. In the case of shorter filaments, their diffuse extension merged with the diffuse side plumes, leaving a dark triangular region in the middle. The formation of Ar/N$_2$ specific diffuse side plumes is likely connected to the presence of long-lived nitrogen species. 

The plasma treatment of polypropylene (PP) in the Ar PSJ discharge was the most uniform. The high treatment uniformity results from the short filament lifetimes and fast jitter speeds caused by the self-organizing nature of filaments trying to maintain an ideal inter-filament distance. Even though a much larger dark region separates the Ar/O$_2$ filaments than in the Ar/N$_2$ discharge, the plasma modification uniformity was better using the Ar/O$_2$ PSJ. The PP reactivity with nitrogen gas species is much lower than with the oxygen ones, and the nitrogen functional groups are less hydrophilic. The nonuniformities observed on the PP treated in the Ar/N$_2$ discharge were removed by increasing the number of passes to two. As published before, in these conditions, the Ar/N$_2$ PSJ was the most efficient to increase PP adhesion to the epoxy glue.

\section{Acknowledgments}
\label{acknowledgements}
This work has been supported by the Czech Science Foundation in the frame of project 20-14105S. CzechNanoLab project LM2018110 funded by MEYS CR is gratefully acknowledged for the financial support of the measurements/sample fabrication at CEITEC Nano Research Infrastructure.

\appendix

%% References


\bibliographystyle{elsarticle-num}

\section*{References}
\bibliography{RF_diagnostics-homogeneity}

\end{document}